\documentclass[english,11pt,nofootinbib,showpacs,notitlepage,usenames,dvipsnames,svgnames,table]{revtex4-1}
\usepackage[T1]{fontenc}
\usepackage{babel}
\usepackage[utf8]{inputenc}
\usepackage{babel,amsmath,amssymb,mathrsfs,amsfonts,calligra,euscript,textcomp,graphicx,wrapfig,fontenc,pst-blur,tabu,lipsum,%
float,empheq,pst-node,environ,framed,ccaption,capt-of,enumitem,subcaption,enumitem,framed,txfonts,lmodern,cancel,diagbox,tabularx,multirow,float}
\usepackage[section]{placeins}
\definecolor{darkgreen}{rgb}{0,.7,0}
\usepackage[colorlinks,linkcolor=blue,citecolor=green,pagebackref=true]{hyperref}
\def\[{\left[}
\def\]{\right]}
\def\({\left(}
\def\){\right)}

\def\1{{\bf CI}}
\def\2{{\bf CII}}
\def\3{{\bf CIII}}

\def\m{\mathcal{M}}

\newcommand{\eq}[1]{\begin{equation}#1\end{equation}}

\newcommand{\const}{\mathop{\rm const}\nolimits}

\newcommand{\bgp}[1]{\bigl(#1\bigr)}
\newcommand{\lrp}[1]{\left(#1\right)}

\newcommand{\nd}{\noindent}

\newenvironment{tightcenter}{%
  \setlength\topsep{0pt}
  \setlength\parskip{0pt}
  \begin{center}
}{%
  \end{center}
}

\everymath{\displaystyle}

\begin{document}

\title{Anisotropic cosmological dynamics in Einstein - Gauss - Bonnet gravity:
an example of dynamical compactification in 7+1 dimensions}

\author{Dmitry Chirkov}
\affiliation{Sternberg Astronomical Institute, Moscow State University, Moscow, Russia}
\affiliation{Bauman Moscow State Technical University, Moscow, Russia}

\author{Alex Giacomini}
\affiliation{Instituto de Ciencias F\'isicas y Matem\'aticas, Universidad Austral de Chile, Valdivia, Chile}

\author{Alexey Toporensky}
\affiliation{Sternberg Astronomical Institute, Moscow State University, Moscow, Russia}
\affiliation{Kazan Federal University, Kremlevskaya 18, Kazan 420008, Russia}

\begin{abstract}
We consider a particular example of dynamical compactification of an anisotropic
7+1 dimensional Universe in Einstein - Gauss - Bonnet gravity. Starting from rather
general totally anisotropic initial conditions a Universe in question evolves towards a product
of two isotropic subspaces. The first subspace expands isotropically, the second
represents an "inner" isotropic subspace with stabilized size. The dynamical evolution
does not require fine-tuning of initial conditions, though it is possible for a particular
range of coupling constants. The corresponding condition have been found analytically
and have been confirmed using numerical integration of equations of motion.
\end{abstract}

\maketitle

\section{Introduction}
Since its proposal in $1915$ General relativity, with its revolutionary idea that gravity is related to space-time geometry, has been an extremely successful theory in describing the universe at large scale.

However General Relativity has still some issues which motivated physicists to explore also alternative theories of gravity. One of the most famous issues is that gravity up to know does not fit in the Standard Model, which successfully describes the other three fundamental interactions. This is due to the fact that the standard model is based on perturbative quantum field theory. However a naive attempt to quantize gravity leads to a non-renormalizable theory. Another famous problem of General Relativity is that in order to fit the astrophysical observations one must assume the existence of dark matter and dark energy. In order to address  issue of quantization of Gravity there have been some proposals which are radical departures from General Relativity like String Theory or Loop Quantum Gravity. Another way to address the mentioned issues of General Relativity, without making so radical departures from the original theory, is to look for so called modified theories of gravity, which replace the action of General relativity by something more general but still maintain the same spirit of gravity as a geometric theory. Actually modified theories of gravity have been proposed already since the early times of General Relativity. Historically the first modification of General Relativity was Kaluza-Klein theory which proposed the existence of an extra compact space dimension (successively more compact extra dimensions were proposed) in order to unify gravity with electromagnetism.

A very popular modified theory of gravity used in order to explain for example the accelerated expansion of the universe is $f(R)$ gravity \cite{f(R)}. Another famous modified theory is Ho\v{r}ava-Lifschitz gravity~\cite{Horava-1,Horava-2,Horava-3} etc.) which has been proposed as quantum theory of gravity and which has also interesting applications in cosmology. Both theories lead to field equations which are higher than second order in derivatives. This causes appearance nonphysical solutions like "false radiation" vacuum isotropic solution~\cite{false_radiation} in $f(R)$.

In the context of higher dimensional gravity, in order to avoid higher order field equations, a very natural choice for a modified theory of gravity is given by Lovelock Gravity \cite{Lovelock}. This theory is the most natural extension of general relativity to higher dimensions in the sense that its action is constructed according to the same principle as the Einstein Hilbert action in four dimensions. Actually in four dimensions it can be shown that the only action which can be built from curvature invariants and which leads to second order field equations is the Einstein Hilbert action plus the Lambda term. Here one should be more precise as one can still add to this action a so called Gauss-Bonnet term $I_{GB}=\int d^4 x \sqrt{-g} \left( R_{\alpha\beta\mu\nu}R^{\alpha\beta\mu\nu} -4R_{\mu\nu}R^{\mu\nu}+R \right)$. This term is topological in four dimensions and does not affect the equations of motion and therefore is usually ignored. However in dimensions higher than four this term does affect the equations of motion but gives only contributions  which are of second order in the derivatives of the metric and therefore there is no good reason to ignore this term. The higher dimensional gravity theory whose action is the sum of the Einstein Hilbert action a cosmological term and the Gauss Bonnet term  is called Einstein-Gauss-Bonnet (EGB) gravity. Actually EGB gravity is only a particular case of a more generic family of gravity theories known as Lovelock Gravity whose action is built from dimensional continuations of topological terms. For example in six dimensions one can add to the EGB action a topological term which is cubic in the curvature. In dimension higher than six this term also affects the equations of motion but its contribution is again only second order in the derivatives. Actually for any even $2N$ dimension one can add to the action a topological term which is of power $N$ in the curvature and whose dimensional continuation to $2N+1$ dimensions give a second order contribution in the derivatives to the equations of motion \cite{Zumino}. For this reason Lovelock gravity can be interpreted as a very natural gravity theory in higher dimensions.

In this paper we will study the dynamic compactification  focusing on EGB Gravity (for some results on compactification scenarios in cubic Lovelock gravity see e.e. \cite{CGTW} and \cite{CGT}). Indeed EGB Gravity as a special case of Lovelock gravity has attracted special interest in gravity physics to the fact that is the simplest case of Lovelock gravity and also because it is the low energy limit of some String theories (see e.g. the discussion in \cite{GG}). Also cosmology, in the context of dynamical compactification, EGB Gravity is the most studied Lovelock Gravity (see e.g.\cite{DFB,DCMTP,MH,MM} and references therein) . A peculiarity of EGB Gravity, being quadratic in the curvature, is that generally there can exist up to two maximally symmetric vacua or none at all. Indeed in EGB Gravity the Lambda term is not directly the cosmological constant as the "cosmological constant" of the maximally symmetric solution is given by a quadratic equation whose coefficients involve all the couplings of the theory \cite{BD}. Depending on the values of the couplings this equation can have up to two solutions or, if the discriminant is negative, no solution at all. Actually the situation where no maximally symmetric solution exist  can happen for all Lovelock theories whose highest term is an even power in the curvature, whereas when the highest term is an odd power in the curvature the situation with no maximally symmetric vacuum does not occur. EGB Gravity in the case where the theory does not admit a maximally symmetric solution  has been, up to now, only very little explored in literatures but it has, as we will see in this paper, a remarkable  importance  in cosmology in the context of dynamical compactification.

In \cite{CGPT1} and \cite{CGPT2} the dynamic compactification to a space-time of the form $\m_4\times \m_D$ in EGB Gravity has been studied, where $\m_4$ is a Lorentzian FRW manifold and $\m_D$ is Euclidean $D$ dimensional compact constant curvature manifold and two independent scale factor. More precisely it was studied under which conditions one has a realistic compactification scenario i.e. a constant Hubble parameter in $\m_4$ and a constant scale factor for $\m_D$. The "inner" space is considered as compact and
negatively curved. Positively curved inner spaces often lead to instability of compactification \cite{Barrow}, however,
in some models (more general than EGB) they can be stable \cite{Rubin}. We leave detailed analysis of possibility of compactification with positively curved inner spaces in EGB gravity to a separate work, assuming here in Sec. 2 and 3 that
the spatial curvature of the inner space is negative as it was done in \cite{CGPT1} and \cite{CGPT2}.
By performing numerical analysis it has been found, that  realistic compactification  scenarios only occur in the region of parameter space where the theory does not admit maximally symmetric solutions. This situation, without a maximally symmetric vacuum  was called in the cited papers  "geometric frustration" \footnote{The term "geometric frustration" is used in statistical mechanics to describe a situation  where it is not possible to reach a state of minimal interaction energy due to a nontrivial topology of the lattice on which the Hamiltonian is defined}. However there was no analytic proof of geometric frustration as a necessary condition for realistic compactification. This was due to the fact that for arbitrary $D$ the equations of motions are too bulky in order find a simple analytic expression. However, due to the fact that in the numerically analysis the  geometric frustration played a crucial role, it is of great interest to find an analytic proof of the necessity of this condition. Therefore, in order to get more manageable equations of motion, it is useful to fix the space-time dimension. This paper we will chose an eight dimensional space-time and study the dynamical compactification with 3 "large" and four "small" space dimensions starting from totally anisotropic configuration. The reason to study the case of $(7+1)$ dimensions in detail is that
it is the lowest dimension case which allows us to get zero effective cosmological constant in large
dimensions subspace when size of inner dimensions is stabilized at some finite and non-zero value \cite{Fr}.
We suppose that the cosmological evolution of 7-dimensional space under consideration consist of two stages: on the first stage the space evolve from a totally anisotropic state to the state with 3-dimensional expanding and 4-dimensional contracting isotropic subspaces; on the second stage constant negative curvature of the $4D$ subspace begins to play the role and provide compactification of these extra dimensions.

Compactification model considered here bases on solutions with exponential time dependence of the scale factor. Such exponential solutions have been initially found in EGB theory in the regime when the Gauss-Bonnet term is dominated~\cite{Ivashchuk-1} and later have been generalized to full EGB theory~\cite{grg10,KPT}. It appeared that these solutions exists only if the space has isotropic subspaces~\cite{CST1,ChPavTop1}. It is important to emphasize that this splitting into isotropic subspaces appears naturally from equations of motion as a condition for such solutions to exist. More detailed study of this situation has revealed that  there are solutions where three dimensions (corresponding to the "real world") are expanding, while remaining dimensions are contracting making the compactification viable.

The structure of the paper is the following: in the next section the EGB action, the compactification ansatz and  equations of motion are presented; in the third section we derive analytically the necessary conditions on the coupling parameters of the action for the compactification regime to exist; the fourth section is devoted to the numerical analysis and in the last section the conclusions will be given.

\section{The set-up}
The Einstein-Gauss-Bonnet action in $(D+1)$-dimensional spacetime $\mathcal{M}$ reads
\eq{S=\int_{\mathcal{M}}d^{D+1}x\sqrt{|g|}\left\{R-\Lambda+\alpha\bgp{R_{\mu\nu\rho\sigma}R^{\mu\nu\rho\sigma}-4R_{\mu\nu}R^{\mu\nu}+R^2}\right\},}
where $R,R_{\alpha\beta},R_{\alpha\beta\gamma\delta}$ are the $(D+1)$-dimensional scalar curvature, Ricci tensor and Riemann tensor respectively; $|g|$ is the determinant of metric tensor; $\Lambda$ is the cosmological term; $\alpha$ is the coupling constant. Here and after we suppose that $\Lambda>0$.

We suppose that the space-time $\mathcal{M}$ is a warped product $\m_4\times \m_D$ where $\m_4$ is a flat Friedman-Robertson-Walker manifold with scale factor $a(t)$, $\m_D$ is a $D$-dimensional Euclidean compact and constant curvature manifold with scale factor $b(t)$ and spatial curvature $\gamma_{D}$ . The ansatz for the metric has the form
\eq{ds^2=-dt^2+a(t)^2d\Sigma^2_{3}+b(t)^2d\Sigma^2_{D}\label{metric}}
Let us denote $H=\frac{a'}{a}$; then $\frac{a''}{a}=H'+H^2$. Taking this into account we get equations of motion \cite{CGPT1, CGPT2}
\eq{\begin{split}
       &\frac{6}{D+1}\left(\frac{2H b'(D+1)!}{b (D-1)!}+\frac{H^2(D+1)!}{D!}+\frac{(\gamma_{D}+b'^2)(D+1)!}{2b^2(D-2)!}+\frac{b''(D+1)!}{b(D-1)!}+\frac{2(H'+H^2)(D+1)!}{D!}\right)+ \\
         &+6D\alpha\Biggl(\frac{(\gamma_{D}+b'^2)^2(D-1)!}{2b^4(D-4)!}+
      \frac{8b''b'H(D-1)!}{b^2(D-2)!}+\frac{4(\gamma_{D}+b'^2)(H'+H^2)(D-1)!}{b^2(D-2)!}+\\
         &\hspace{1.5cm} +\frac{4H^2b''}{b}+\frac{4Hb'(\gamma_{D}+b'^2)(D-1)!}{b^3(D-3)!}+\frac{4H^2b'^2(D-1)!}{b^2(D-2)!}+\\
         &\hspace{1.5cm} +\frac{2H^2(\gamma_{D}+b'^2)(D-1)!}{b^2(D-2)!}+\frac{8(H'+H^2)Hb'}{b}+\frac{2b''(\gamma_{D}+b'^2)(D-1)!}{b^3(D-3)!}\Biggr)-3\Lambda=0
    \end{split}
\label{E1}}
\eq{\begin{split}
       &\frac{6}{D+1}\left(\frac{H b'(D+1)!}{b (D-2)!}+\frac{H^2(D+1)!}{(D-1)!}+\frac{(\gamma_{D}+b'^2)(D+1)!}{6b^2(D-3)!}+\frac{b''(D+1)!}{3b(D-2)!}+\frac{(H'+H^2)(D+1)!}{(D-1)!}\right)+ \\
         &+6D\alpha\Biggl(\frac{(\gamma_{D}+b'^2)^2(D-1)!}{6b^4(D-5)!}+\frac{2H^2(\gamma_{D}+b'^2)(D-1)!}{b^2(D-3)!}+
      \frac{4b''b'H(D-1)!}{b^2(D-3)!}+\frac{4H^3b'(D-1)!}{b(D-2)!}+\\
         & \hspace{1.5cm}+\frac{4H^2b''(D-1)!}{b(D-2)!}+\frac{2Hb'(\gamma_{D}+b'^2)(D-1)!}{b^3(D-4)!}+\frac{4H^2b'^2(D-1)!}{b^2(D-3)!}+
         \frac{8(H'+H^2)Hb'(D-1)!}{b(D-2)!}+ \\
         &\hspace{1.5cm}+\frac{2b''(\gamma_{D}+b'^2)(D-1)!}{3b^3(D-4)!}+\frac{2(H'+H^2)(\gamma_{D}+b'^2)(D-1)!}{b^2(D-3)!}+4H^2(H'+H^2)\Biggr)-D\Lambda=0
    \end{split}
\label{E2}}
and constraint
\eq{\begin{split}
       &\frac{6}{D+1}\left(\frac{H b'(D+1)!}{b (D-1)!}+\frac{H^2(D+1)!}{D!}+\frac{(\gamma_{D}+b'^2)(D+1)!}{6b^2(D-2)!}\right)+
         6D\alpha\Biggl(\frac{(\gamma_{D}+b'^2)^2(D-1)!}{6b^4(D-4)!}+\\
      &\hspace{1.5cm}+\frac{2H^2(\gamma_{D}+b'^2)(D-1)!}{b^2(D-2)!}+\frac{4H^3b'}{b}+\frac{4H^2b'^2(D-1)!}{b^2(D-2)!}+
      \frac{2Hb'(\gamma_{D}+b'^2)(D-1)!}{b^3(D-3)!}\Biggr)-\Lambda=0.
    \end{split}
\label{E0}}
 In order to get a viable compactification scenario we focus on the case $D=4,\;\gamma_{D}=-1$.

\section{Conditions for existence of a realistic compactification regime \label{cond-for-stab}}
Let us find out for what values of $\alpha$ and $\Lambda$ compactification regime does exist and compare them with the corresponding values of $\alpha$ and $\Lambda$ for which isotropic solution exists.

A phenomenologically realistic compactification scenario suggests that at late times the "large dimensions" have an accelerated expansion whereas the extra dimensions tend to a constant size namely  $H'(t),b'(t),b''(t)\rightarrow0,\;H(t)\rightarrow H_0,\;b(t)\rightarrow b_0$ as $t\rightarrow\infty$; $b_0,H_0=\const$. For each dimension we substitute the asymptotic conditions $H'(t)=b'(t)=b''(t)=0$ into the equations of motion, get the asymptotic equations and find out for which $\alpha$ and $\Lambda$ these equations have solutions.

Let us denote $\xi=\alpha\Lambda$. Taking into account that $\Lambda>0$ we conclude that both $\alpha$ and $\xi$ has the same signs.

Asymptotic equations for (7+1)-dimensional EGB cosmological models are
\eq{24\alpha^2b_0^2H_0^4+(12\alpha b_0^2-144\alpha^2)H_0^2-\xi b_0^2-6\alpha=0}
\eq{(6\alpha b_0^4-144\alpha^2b_0^2)H_0^2-\xi b_0^4-12\alpha b_0^2+24\alpha^2=0}
where $\xi=\alpha\Lambda$. Solving these equations we obtain:
\begin{enumerate}
  \item $H_0^2=-\frac{1}{b_0^2},\;b_0^2=\frac{\alpha(-9\pm\sqrt{81+168\xi})}{\xi},\;\xi\geqslant-\frac{27}{56},\;\xi\ne0\;-$ there are no real-valued solutions in this case.
  \item $b_0^2=\frac{2\alpha(15+\sqrt{45-120\xi})}{2\xi+3}>0\Longleftrightarrow\xi\in\lrp{-\infty;-\frac{3}{2}}\cup\Bigl(0;\frac{3}{8}\Bigr]$,\\
      $H_0^2=\frac{(-7\xi-3)\sqrt{45-120\xi}+24\xi^2-63\xi-36}{\alpha(15+\sqrt{45-120\xi})(21+24\xi-\sqrt{45-120\xi})}>0\Longleftrightarrow\xi<0,\,\xi\ne-\frac{11}{24}.$\vspace{0.2cm}\\
      Real-valued $b_0$ and $H_0$ exist simultaneously for $\xi<-\frac{3}{2}$.
  \item $b_0^2=\frac{2\alpha(15-\sqrt{45-120\xi})}{2\xi+3}>0\Longleftrightarrow\xi\in\Bigl(0;\frac{3}{8}\Bigr],$\\
      $H_0^2=\frac{(-7\xi-3)\sqrt{45-120\xi}-24\xi^2+63\xi+36}{\alpha(-15+\sqrt{45-120\xi})(21+24\xi+\sqrt{45-120\xi})}>0\Longleftrightarrow\xi\in\lrp{-\frac{3}{2};0}$\vspace{0.2cm}\\
      Real-valued $b_0$ and $H_0$ do not exist simultaneously.
\end{enumerate}
We conclude that compactified solutions exist only if $\alpha\Lambda<-\frac{3}{2}$ ($\alpha<0;\Lambda>0$). This condition covers the condition for non existence of a maximally symmetric solution (i.e. geometric frustration). We remind a reader that absence of compactification in the situation when an isotropic solution exists have been already confirmed by the numerical calculations for different numbers of inner dimensions. It is easy to check that the isotropic regime is solution to the equation
\eq{840\alpha H^4+42H^2-\Lambda=0}
This equation has at least one real solution if $-\frac{21}{40}\leqslant\alpha\Lambda<0,\;\alpha<0$ or $\alpha\Lambda\geqslant0,\;\alpha\in\mathbb{R}$ and one can see that compactification regime can not coexists with isotropic regime for the $(7+1)$ dimension case. It is worth to point out that the condition of geometric frustration defines an open set of the parameter space and therefore does not constitute a fine tuning.

\section{Numerical calculations}

The scenario we are studying includes 2 stages: 1) evolution of totally anisotropic flat $7D$ space to a warped product of $3D$ expanding and $4D$ contracting isotropic subspaces; 2) compactification due to the curvature of $4D$ subspace.

\subsection{Stage 1: splitting of flat space onto isotropic subspaces \label{stage-1}}

It is proved~\cite{ChPavTop1} that non constant-volume\footnote{For definition of constant-volume exponential solutions see~\cite{ChPavTop2}.} exponential solutions in Lovelock gravity exist only if the space has isotropic subspaces. As only configurations with 3-dimensional isotropic subspaces has physical meaning, the question arises: can such configuration be a natural outcome of the dynamical evolution?

It have been shown that in EGB gravity it can be no more than 3 different Hubble parameters independently of the number of dimensions~\cite{Iv-16}, so the maximum number of isotropic subspaces in EGB model is three. Numerical study of (5+1)- and (6+1)-dimensional flat EGB models showed~\cite{CT:splitting} that the nonsingular cosmological evolution in most cases leads from initially totally anisotropic stage to a warped product of two isotropic subspaces; particularly, $3D$ isotropic subspace appears even if we start from quite arbitrary (anisotropic) initial values of the Hubble parameters (see tables III and VI in~\cite{CT:splitting}). Numerical analysis carried out in this work shows that the same is true for the (7+1)-dimensional model.

On the first stage we consider flat anisotropic $(7+1)$-dimensional spacetime with the metric ansatz  of the form
\eq{ds^2=-dt^2+\sum\limits_{k=1}^D a^2_k(t)dx_k^2}
where in general $a_1(t)\ne a_2(t)\ne \ldots\ne a_D(t)$ are scale factors. In spatially flat model scale factors $a_k(t)$ are defined only up to a constant factor, so we use the Hubble parameters $H_k=\frac{\dot{a}_k}{a_k},\;k=1,\ldots,D$ instead of scale factors. Equations of motion and constraint read
\eq{\begin{split}
      2\sum\limits_{i\ne j}(\dot{H}_i+H_i^2)&+2\sum\limits_{\{i>k\}\ne j}H_i H_k+ \\
        & +8\alpha\sum\limits_{i\ne j}(\dot{H}_i+H_i^2)\sum\limits_{\{k>l\}\ne\{i, j\}}H_k H_l+24\alpha\sum\limits_{\{i>k>l>m\}\ne j}H_i H_k H_l H_m=\Lambda,\quad j=1,\ldots,D
    \end{split}\label{eq.of.motion}}
\eq{2\sum\limits_{i>j}H_i H_j+24\alpha\sum\limits_{i>j>k>l}H_i H_j H_k H_l=\Lambda\label{constraint}\,,}
These equations we solve numerically. We chose randomly 10 pairs $(\alpha;\Lambda)$ such that $\alpha\Lambda<-\frac{3}{2}$ (necessary condition for compactified solutions to exist); for each pair $(\alpha;\Lambda)$ we chose randomly 1000 sets of initial values of the Hubble parameters with different number of positive and negative elements in a set; we denote a set of initial values with $p$ positive and $q$ negative elements by $(p-q)$. Table~\ref{D=7_alpha=3.5821_lambda=-1.7354} demonstrates typical distribution of solutions by types depending on the number of positive and negative initial values of the Hubble parameters for $\alpha=-3.5821,\,\Lambda=1.7354$; distributions obtained for other 9 pairs $(\alpha;\Lambda)$ differ from this by only 2-3\%. Following the notations used in~\cite{CT:splitting}, we denote solution with $k$-dimensional expanding subspace and $m$-dimensional contracting subspace by $\{k,-m\}$.

\begin{table}[!h]
\caption{The percentage ratio of solutions of different classes for $\alpha=-3.5821,\,\Lambda=1.7354$}
\label{D=7_alpha=3.5821_lambda=-1.7354}
\centering
  \begin{tabular}{|c|c|c|c|c|}
  \hline
  \diagbox[width=10em]{Initial\\configuration}{Resulting\\solution} & \{3,-4\} & \{4,-3\} & \{5,-2\} & non-standard singularity  \\
  \hline
  (1-6) & \multicolumn{4}{|c|}{No Hubble parameters with positive sum have been found} \\
  \hline
  (2-5) & 0 & 0 & 0 & 100 \\
  \hline
  (3-4) & 28 & 20 & 44 & 8 \\
  \hline
  (4-3) & 3 & 43 & 53 & 1 \\
  \hline
  (5-2) & 0 & 0 & 66 & 34 \\
  \hline
  (6-1) & 0 & 0 & 0 & 100 \\
  \hline
  \end{tabular}
\end{table}

It should be emphasized that all initial values in a set are different, i.e. we start from a totally anisotropic configuration. Since we are interesting only in stable solutions we require that the sum of initial values of the Hubble parameters is positive~\cite{Pavl-15,ErIvKob-16,ChT}.

Numerical calculations show that among solutions with $3D$ isotropic subspace there are solutions with $3D$ expanding and $4D$ contracting isotropic subspaces (we denote them by $\{3,-4\}$) as well as solutions with $4D$ expanding and $3D$ contracting isotropic (we denote them by $\{4,-3\}$).

It should be noted that solutions which are the warped product of 3D and 4D isotropic subspaces (of all possible types: ($\{3,-4\}$, $\{4,-3\}$, $\{-4,-3\}$ and $\{4,3\}$)) exist only if $\alpha\Lambda<0$. Indeed, such solutions obey the equations~\cite{ChPavTop1}
\eq{3h^2+6hH+H^2=-\frac{1}{4\alpha},\quad 5h^4+12h^3H+11h^2H^2+6hH^3+H^4=-\frac{\Lambda}{24\alpha}}
Numerical study (for wide enough range of $\alpha$ and $\Lambda$) shows that these equations has solutions only for $\alpha\Lambda<0$; since we assume that $\Lambda>0$ it means that $\alpha$ must be negative.

We should note that when we are solving (for negative $\alpha$) equations~(\ref{eq.of.motion})-(\ref{constraint}), the resulting branch of solution ($\{3,-4\}$ or $\{4,-3\}$) which we reach during cosmological evolution depends only on initial values of the Hubble parameters and does not depend on values of $\alpha,\Lambda$.

So, we have convinced that the cosmological evolution can lead from a totally anisotropic stage to the stage with $3D$ expanding and $4D$ contracting isotropic subspaces, so that the form of metric~(\ref{metric}) is now is not put "by hand" but represents a result of  dynamical evolution from a totally anisotropic space. Starting from the metric~(\ref{metric}) we can now proceed to consider stabilization of $4D$ contracting isotropic subspace.
\pagebreak
\subsection{Stage 2: compactification}
\begin{wrapfigure}{r}{250pt}
\includegraphics[width=250pt,height=250pt,keepaspectratio]{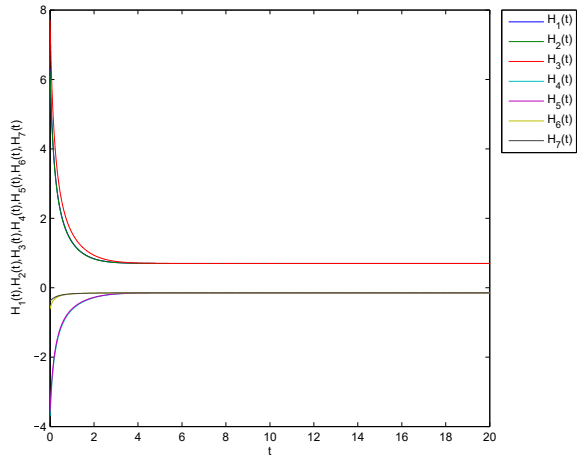}
\caption{\footnotesize Splitting of space into 3D expanding and 4D contracting isotropic subspaces.}
\label{splitting}
\end{wrapfigure}
We conjecture that curvature of space is negligible at the beginning of the cosmological evolution. Let us assume that, in the process of the evolution, the space was split on $3D$ expanding isotropic subspace and 4D contracting isotropic subspace before
the curvature starts to be important, so we can use the results of previous subsection. The scale factor $a(t)$ of the 3D subspace is increasing; the scale factor $b(t)$ of the $4D$ subspace is decreasing. This means that while the curvature of the $3D$ subspace continues to decrease, the curvature of the $4D$ subspace increases and causes stabilization of extra dimensions. This is confirmed by numerical calculations.

There are two possible outcomes of the evolution: singular solution and compactified solution with $H(t)\rightarrow\const,\,\frac{b'(t)}{b(t)}\rightarrow0,\,b(t)\rightarrow\const$.
We perform the numerical analysis as follows: once the dynamical equations without curvature is solved and $\{3,-4\}$-solutions is obtained, we solve dynamical equations with spatial curvature of $4D$ isotropic subspace. "Turning on" curvature causes appearance of terms proportional to $b_0^{-2},\;b_0^{-3}$ and $b_0^{-4}$ in the equations of motion, so solving
dynamical equations with curvature we can not start exactly from a $\{3,-4\}$-solution obtained on the previous step -- initial values of the Hubble parameters must distinguish from what we already have at least by value of $~b_0^{-2}$.

Numerical calculations performed for the same 10 pairs $(\alpha;\Lambda)$ which we used for exploring first stage of our scenario (see subsection~\ref{stage-1}) show that we obtain only compactified solutions until initial values of the Hubble parameters deviate from flat $\{3,-4\}$-solutions by a value $\lesssim 10^4\cdot b_0^{-2}$; otherwise, singular solutions begin to appear.

\begin{figure}[!h]
\begin{minipage}[h]{.4\linewidth}
\center{\includegraphics[width=\linewidth]{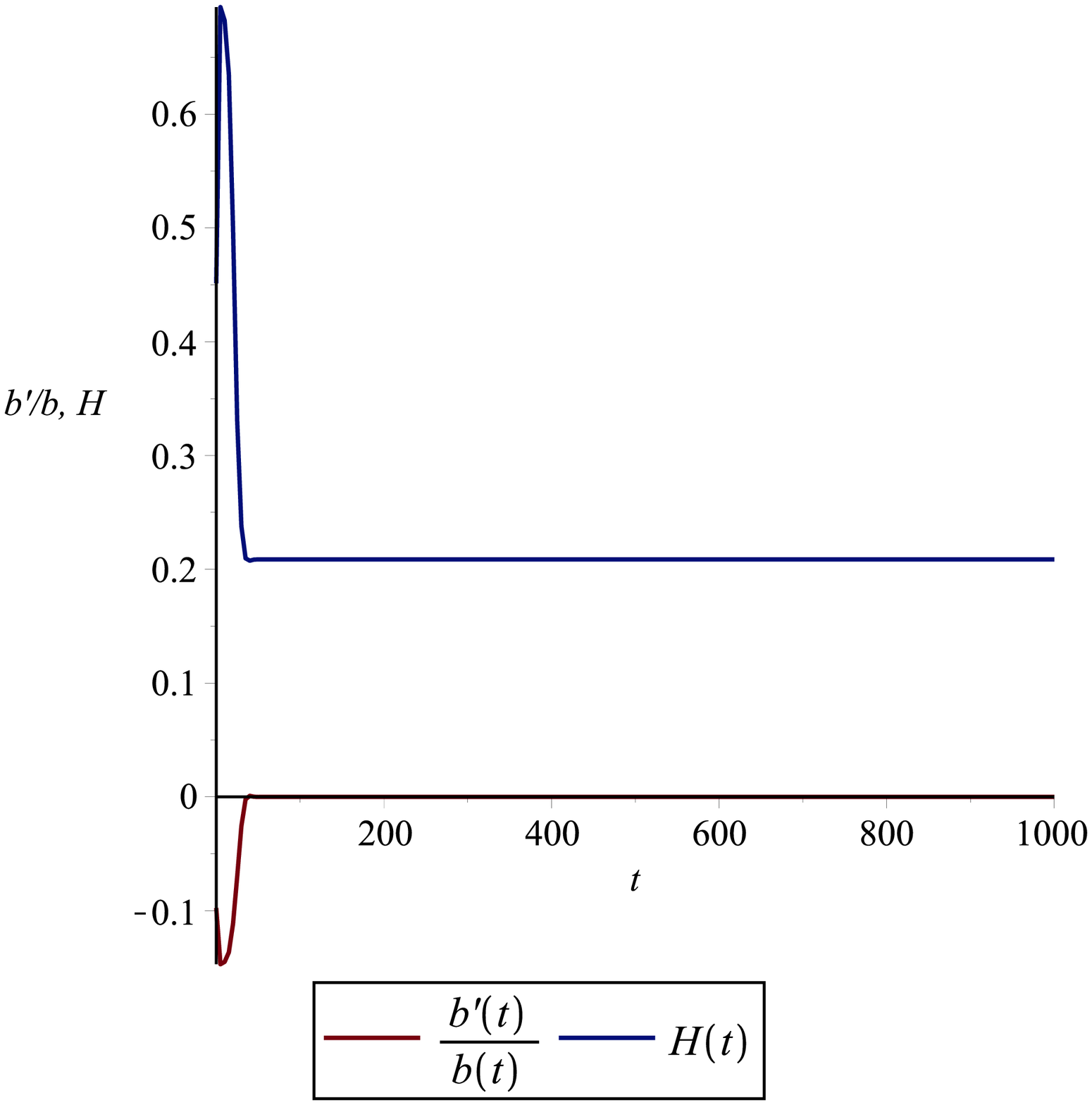} \\ a)}
\end{minipage}
\hfill
\begin{minipage}[h]{.4\linewidth}
\center{\includegraphics[width=\linewidth]{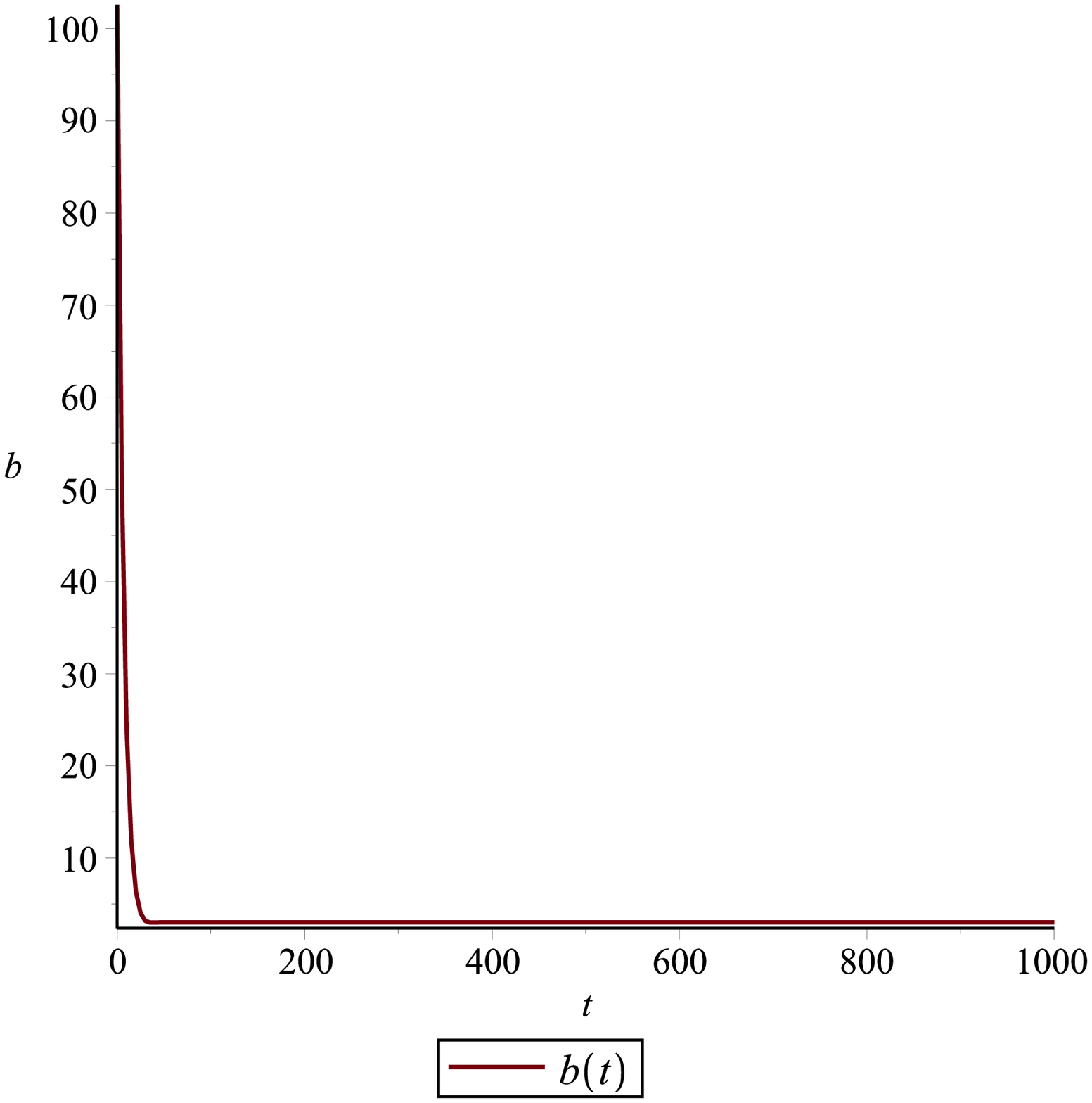} \\ b)}
\end{minipage}
\caption{\footnotesize Compactification regime: $H(t)\rightarrow0.21,\,\frac{b'(t)}{b(t)}\rightarrow0,\,b(t)\rightarrow3.02$.}
\label{evolution}
\end{figure}

 We remind  a reader that compactification regime exists only if $\alpha\Lambda<-\frac{3}{2}$ (see Sec.~\ref{cond-for-stab}), our numerical results indicate that if we start from the vicinity of appropriate  exponential solution, this condition is not only necessary, but sufficient. Examples of numerical solutions are presented in Figs.~\ref{splitting} and~\ref{evolution}. Fig.~\ref{splitting} illustrates the process of splitting of space into $3D$ expanding and $4D$ contracting isotropic subspaces; Fig.~\ref{evolution} illustrates the process of compactification.

\section{Conclusions}
In the present paper we have considered cosmological dynamics of
(7+1) dimensional anisotropic Universe in Einstein-Gauss-Bonnet gravity
both analytically and numerically. Our analytical study considers conditions for compactification scenario.
In this scenario the outcome of Universe evolution is exponentially expanding (with the constant Hubble rate $H$) large dimensions
and stabilized (with the constant size $b$) small dimensions. After substituting $H=\const$ and $b=\const$ into the equations of motion~(\ref{E1})-(\ref{E0}) we get two independent algebraic equations for $H$ and $b$.
Compactified solutions exist only for $\alpha \Lambda \le -3/2$ (Sec.~\ref{cond-for-stab}). Since this interval excludes the possibility for existing of isotropic solution, we have found an explanation why the compactification have never seen for the range of parameters admitting an isotropic solution.

For numerical study we have chosen a set of 10 different combinations of coupling constants, all of them satisfy the above condition
for compactification. We have considered two different regimes:
\begin{itemize}
\item {Evolution of a general flat anisotropic Universe according to equations of motion~(\ref{eq.of.motion})-(\ref{constraint}).}
\item {Evolution of a warped product~(\ref{metric}) with the negatively curved inner dimensions according to equations~(\ref{E1})-(\ref{E0}).}
\end{itemize}

We have found that a large set of initially totally anisotropic metrics ends up with the warped product of two isotropic subspaces.
This allows us to start with them during the second regime, where we take into account spatial curvature of the inner space.
In particular, we have studied trajectories reaching at the first stage the warped product of three expanding and four contracting
dimensions. The initial conditions for the second stage have been chosen near the outcome of the first stage (we can not simply equate
the result of the first stage with the initial conditions for the second stage due to presence of spatial curvature neglected at the
first stage).  Our numerical results show that all  trajectories chosen this way lead to stabilization of the inner dimensions.
This means that the model we have studied admits successful dynamical compactification without fine-tuning.

It is also worth to point out that in cubic Lovelock theory there is no possibility to achieve geometric frustration which leads to the conjecture
that dynamical compactification behaves very differently for Lovelock Gravities whose highest curvature term is an even or odd power.
This will be a subject of future investigation.

\appendix

\section{Geometric frustration for low-dimensional cases}
In this section we prove that compactification and isotropization regimes do not coexist in (5+1) and (6+1)-dimensional EGB cosmological models. For (5+1)-dimensional model we also find all possible values of $\xi=\alpha\Lambda$ for which compactified solutions exist; it is much more complicated to find all such $\xi$ for (6+1)-dimensional model, so in the latter case we restrict ourselves by proving that compactification regime can not coexist with isotropic regime.
\subsection{(5+1)-dimensional model}
\nd Asymptotic equations in (5+1)-dimensional EGB cosmological model:
\eq{b_0^{2}(\xi-6H_0^{2}\alpha)+24H_0^{2}{\alpha}^{2}+2\alpha=0\label{constr-bh-5+1}}
\eq{24H_0^{4}\alpha^{2}+12H_0^{2}\alpha-\xi=0\label{eq-bh-5+1}}
Let $z=H_0^{2};\;y=b_0^{2}$. Then equations~(\ref{constr-bh})-(\ref{eq-bh}) take the form:
\eq{y(\xi-6z\alpha)+24z{\alpha}^{2}+2\alpha=0\Longleftrightarrow y=\frac{2\alpha\left(12\,z\alpha+1\right)}{6\,z\alpha-\xi}\label{constr-yz}}
\eq{24z^2\alpha^{2}+12z\alpha-\xi=0\Longleftrightarrow z_{\pm}=\frac{-1\pm\sqrt{1+\frac{2}{3}\xi}}{4\alpha}\label{eq-yz}}
Substituting $z_+$ and $z_-$ into expression for $y$ (see~(\ref{constr-yz})) we obtain correspondingly
\eq{y_+=\frac{4\alpha\left(\sqrt{1+\frac{2}{3}\xi}-\frac{2}{3}\right)}{\sqrt{1+\frac{2}{3}\xi}-\lrp{1+\frac{2}{3}\xi}}\quad\mbox{and}\quad
y_-=\frac{4\alpha\left(\sqrt{1+\frac{2}{3}\xi}+\frac{2}{3}\right)}{\sqrt{1+\frac{2}{3}\xi}+\lrp{1+\frac{2}{3}\xi}}}
One can see that real values of $z_{\pm}$ and $y_{\pm}$ exist only for $\xi>-\frac{3}{2}$. Also both $z$ and $y$ must be positive. Table~\ref{5+1} demonstrates values of $\xi$ for which positive $z$ and $y$ exist. This allows us to conclude that compactification in (5+1) model is possible iff $\xi\in\lrp{-\frac{3}{2};-1}$.

\begin{table}[!h]
  \centering\everymath{\displaystyle}
  \caption{Values of $\xi$ for which positive $z$ and $y$ exist}\label{5+1}
  \begin{tabular}{|c|c|c|}
    \hline
     & $\alpha>0$ & $\alpha<0$ \\
    \hline
    $z_+$ & $\xi\in[0;+\infty)$ & $\xi\in\left(-\frac{3}{2};-1\right)$ \\
    \hline
    $y_+$ & there are no positive $\xi$ & $\xi\in\left(-\frac{3}{2};-\frac{5}{6}\right)$ \\
    \hline\hline
    $z_-$ & $\xi\in\emptyset$ & $\xi\in\left(-\frac{3}{2};0\right]$ \\
    \hline
    $y_-$ & $\xi\in[0;+\infty)$ & $\xi\in\emptyset$ \\
    \hline
  \end{tabular}
\end{table}

Isotropic solutions in (5+1)-dimensional EGB cosmological model are defined as
\eq{H^2=\frac{-1\pm\sqrt{1+\frac{6}{5}\xi}}{12\alpha}}
For $\xi\in\Bigl[-\frac{5}{6};0\Bigr)$ we have 4 possible solutions:
\eq{H=\pm\sqrt{\frac{-1\pm\sqrt{1+\frac{6}{5}\xi}}{12\alpha}},\;\alpha<0}
For $\xi>0$ we have 2 possible solutions:
\eq{H=\pm\sqrt{\frac{-1+\sqrt{1+\frac{6}{5}\xi}}{12\alpha}},\;\alpha>0}
Taking into account that isotropic solutions exist for $\xi\geqslant-\frac{5}{6}$, but asymptotic equations~(\ref{constr-bh-5+1})-(\ref{eq-bh-5+1}) has no real solutions for $\xi\geqslant-\frac{5}{6}$, we conclude isotropization and compactification regimes do not coexist in (5+1)-dimensional EGB cosmological model.

Note that numerical attempts to find compactification solution in $(5+1)$ dimensions have been unsuccessful \cite{Pavl}, and the
reason for that is still unknown.

\subsection{(6+1)-dimensional model}
\nd Asymptotic equations in (6+1)-dimensional EGB cosmological model:
\eq{\left(6\alpha b_0^4-72\alpha^{2}b_0^{2}\right)H_0^{2}-\xi b_0^{4}-6\alpha b_0^{2}=0\label{constr-bh}}
\eq{24H_0^{4}\alpha^{2}b_0^{4}+\left(12\alpha b_0^{4}-48\alpha^{2}b_0^{2}\right)H_0^{2}-\xi b_0^{4}-2\alpha b_0^{2}=0\label{eq-bh}}
Let $z=H_0^{2};\;y=b_0^{2}$. Then equations~(\ref{constr-bh})-(\ref{eq-bh}) take the form:
\eq{\left(6\alpha y^2-72\alpha^{2}y\right)z-\xi y^2-6\alpha y=0\label{constr-yz}}
\eq{24\alpha^{2}z^2y^2+\left(12\alpha y^2-48\alpha^{2}y\right)z-\xi y^2-2\alpha y=0\label{eq-yz}}
Note that both $z$ and $y$ must be positive. Solving~(\ref{constr-yz}) to $z$ we get
\eq{z=\frac{y\xi+6\alpha}{6\alpha(y-12\alpha)}\label{z}}
Substitution~(\ref{z}) into~(\ref{eq-yz}) gives us equation for $y$:
\eq{\xi\left(\xi+\frac{3}{2}\right){y}^{3}+15\,\alpha\,{y}^{2}-72\left(\xi+\frac{5}{2}\right){\alpha}^{2}y+432{\alpha}^{3}=0\label{y}}

 First,  we show that there are no real-valued solutions of the equations~(\ref{constr-bh})-(\ref{eq-bh}) for $\xi>-\frac{3}{2}$.
 It is easy to check that
\eq{F'(y)=3\xi\lrp{\xi+\frac{3}{2}}y^2+30\alpha y-72\alpha^2\lrp{\xi+\frac{5}{2}}}
\eq{F'(y)=0\Longleftrightarrow y_{\pm}=\frac{-5\alpha\pm5|\alpha|\sqrt{1+\frac{24}{25}\xi\lrp{\xi+\frac{3}{2}}\lrp{\xi+\frac{5}{2}}}}{\xi\lrp{\xi+\frac{3}{2}}}}
\textbf{I.} $-\frac{3}{2}<\xi<0\Longrightarrow\alpha<0;\;\xi\lrp{\xi+\frac{3}{2}}<0;\;y_+<0,\,y_-<0$, therefore $F'(y)<0$ for all $y>0$, so function $F(y)$ decrease monotonically for $y>0$. Taking into account that $F(0)=432{\alpha}^{3}<0$ and $F(y)\xrightarrow[y\rightarrow+\infty]{}-\infty$ we conclude that function $F(y)$ has no positive roots for $\xi\in\Bigl(-\frac{3}{2};0\Bigr)$ and, therefore, there does not exist real values of $b_0$ for $\xi\in\Bigl(-\frac{3}{2};0\Bigr)$.

\nd \textbf{II.} $\xi>0\Longrightarrow\alpha>0;\;\xi\lrp{\xi+\frac{3}{2}}>0;\;y_-<0,\,y_+>0$. Function $F(y)$ has  minimum at $y_+>0$; $F(0)=432{\alpha}^{3}>0,\;F(y)\xrightarrow[y\rightarrow+\infty]{}+\infty$, so if $F(y_+)<0$ then $F(y)$ has 2 positive zeros.
Substituting $y_+$ into function $F$ we get

\eq{\hspace{-0.5cm}F(y_+)=-\frac{2\alpha^3\lrp{125\lrp{Q^{3/2}-1}-18\xi\left(\xi+\frac{3}{2}\right)(12\,{\xi}^{2}+28\,\xi+25)}}{\xi^2\lrp{\xi+\frac{3}{2}}^2},\;
Q=1+\frac{24}{25}\xi\lrp{\xi+\frac{3}{2}}\lrp{\xi+\frac{5}{2}}}
and it is easy to see that it is negative for any $\xi>0$, $\alpha>0$.


Once we find $y$ from~(\ref{y}), we should substitute it to~(\ref{z}) and find $z$; since $z$ must be positive, solution of~(\ref{y}) must obey inequality $y>12\alpha$.

Since $F(12\alpha)=1728\alpha^3\lrp{\xi+\frac{1}{2}}^2>0$ the point $y=12\alpha$ can not been located between two roots of the
equation $F(y)=0$. Thus, if $y_+<12\alpha$ then both roots are smaller than $12\alpha$.

\nd In order to compare $y_+$ and $12\alpha$ we introduce function $G(\xi)\equiv\frac{y_+(\xi,\alpha)}{12\alpha}=
\frac{5\lrp{\sqrt{1+\frac{24}{25}\xi\lrp{\xi+\frac{3}{2}}\lrp{\xi+\frac{5}{2}}}-1}}{12\xi\lrp{\xi+\frac{3}{2}}}$. It is easy to check that $G'(\xi)<0$ for $\xi>0$, so function $G(\xi)$ decrease monotonically for all positive values $\xi$. On the other hand $\lim\limits_{\xi\rightarrow0}G(\xi)=\frac{1}{2}$. It means that $G(\xi)<\frac{1}{2}$ for $\xi>0$ and, therefore, $y_+<6\alpha$. So, all positive roots of function $F(y)$ are smaller than $12\alpha$, values of $z$ corresponding to these roots are negative and, therefore, there does not exist real values of $H_0$ for $\xi>0$.

Second, we turn to isotropic solutions in (6+1)-dimensional EGB cosmological model:
\eq{H^2=\frac{-1\pm\sqrt{1+\frac{8}{5}\xi}}{24\alpha}}
For $\xi\in\Bigl[-\frac{5}{8};0\Bigr)$ we have 4 possible solutions:
\eq{H=\pm\sqrt{\frac{-1\pm\sqrt{1+\frac{8}{5}\xi}}{24\alpha}},\;\alpha<0}
For $\xi>0$ we have 2 possible solutions:
\eq{H=\pm\sqrt{\frac{-1+\sqrt{1+\frac{8}{5}\xi}}{24\alpha}},\;\alpha>0}

So, isotropic solutions exist for $\xi\geqslant-\frac{5}{8}$. Remember the  asymptotic equations~(\ref{constr-bh})-(\ref{eq-bh}) has no real solutions for $\xi\geqslant-\frac{3}{2}$. Consequently, isotropization and compactification regimes do not coexist in (6+1)-dimensional EGB cosmological model.

Compactification regime in $(6+1)$ dimensions have been found numerically \cite{Pavl}. Note, however, that structure of equations
prevents solution with zero effective cosmological constant in the bigger subspace since $H=0$ requires $b=0$.

\section*{Acknowledgements}

The work of A. V. T.  was supported by
 RFBR grant 20-02-00411 and
by the Russian Government Program of Competitive Growth of Kazan Federal University.
The work of A. G. was supported by FONDECYT grant n. 1200293.

\end{document}